# *In silico* design, *in vitro* construction and *in vivo* application of synthetic small regulatory RNAs in bacteria

Michel Brück[1,2], Bork A. Berghoff[2], Daniel Schindler[1,3,*]

[1] Max Planck Institute for Terrestrial Microbiology, Karl-von-Frisch-Str. 10, D-35043 Marburg, Germany

[2] Institute for Microbiology and Molecular Biology, Justus-Liebig University Giessen, 35392 Giessen, Germany

[3] Center for Synthetic Microbiology (SYNMIKRO), Philipps-University Marburg, Karl-von-Frisch-Str. 14, D-35043 Marburg, Germany

* Correspondence to: daniel.schindler@mpi-marburg.mpg.de

**Abstract**

Small regulatory RNAs (sRNAs) are short non-coding RNAs in bacteria capable of post-transcriptional regulation. sRNAs have recently gained attention as tools in basic and applied sciences for example to fine-tune genetic circuits or biotechnological processes. Even though sRNAs often have a rather simple and modular structure, the design of functional synthetic sRNAs is not necessarily trivial. This protocol outlines how to use computational predictions and synthetic biology approaches to design, construct and validate synthetic sRNA functionality for their application in bacteria. The computational tool, SEEDling, matches the optimal seed region with the user-selected sRNA scaffold for repression of target mRNAs. The synthetic sRNAs are assembled using Golden Gate cloning and their functionality is subsequently validated. The protocol uses the *acrA* mRNA as an exemplary proof-of-concept target in *Escherichia coli*. Since AcrA is part of a multidrug efflux pump, *acrA* repression can be revealed by assessing oxacillin susceptibility in a phenotypic screen. However, in case target repression does not result in a screenable phenotype, an alternative validation of synthetic sRNA functionality based on a fluorescence reporter is described.

**Key words:** synthetic sRNAs, Golden Gate cloning, post-transcriptional regulation, synthetic biology, synthetic sRNA prediction, functional sRNA profiling, seed region

## 1 Introduction

Synthetic biology aims to see biological functions and concepts from the perspective of an engineer. Synthetic biologists perform characterization and standardization of molecular functions and try to apply them towards a greater aim, for example the construction and improvement of producer strains. Producer strains often contain a heterologous pathway which has to be optimized on multiple levels. The toolbox for gene expression is growing constantly and many tools for the precise control of gene expression by transcription factors are available. In recent years, tools for post-transcriptional regulation gained increasing interest. Post-transcriptional regulation can be achieved, e.g., by CRISPR/Cas systems. They rely on the interplay of a large protein (Cas13 or Cas7-11) and co-expression of a respective guide RNA [1-5]. However, CRISPR/Cas systems for post-transcriptional regulation may interfere with the cellular machinery in bacteria because the mRNA-targeting CRISPR/Cas systems may induce dormancy or cell death [3]. Other elements for post-transcriptional control are riboswitches which are integrated into the 5´ untranslated region (UTR) of mRNAs for precise control of translation [6]. However, riboswitches need the genetic engineering of every target gene which may be cumbersome in particular in non-model organisms.

Bacteria exploit manifold post-transcriptional regulators to quickly adapt to environmental stimuli. Among them are small regulatory RNAs (sRNAs) which are expressed in specific growth phases or upon environmental stress conditions [7]. Expression of natural sRNAs often prevents translation of mRNAs by an antisense mechanism blocking translation initiation (Fig. 1A). The RNA chaperone Hfq is an important factor facilitating many sRNA/mRNA interactions. Furthermore, Hfq may recruit RNase E and mediate RNase E-dependent mRNA decay [8,9]. sRNAs, like the well-studied RybB sRNA, have a modular structure consisting of a seed region and a scaffold (Fig. 1B). The scaffold usually provides a Rho-independent transcriptional terminator (hairpin loop followed by consecutive uracils) and mediates interaction with Hfq. The seed region is capable of binding the corresponding target mRNAs by base-pairing. Binding of the seed region within the translation initiation region (TIR) of an mRNA usually prevents protein translation. Because of their simple and modular structure, sRNAs are gaining an increased interest for synthetic biology and biotechnological applications. The rational design of seed regions together with suitable scaffolds supports engineering of synthetic sRNAs, which can subsequently be applied for targeted repression of mRNA translation. The modularity and small size allow fast and large-scale synthesis of multiple variants and parallel testing of the resulting synthetic sRNAs.

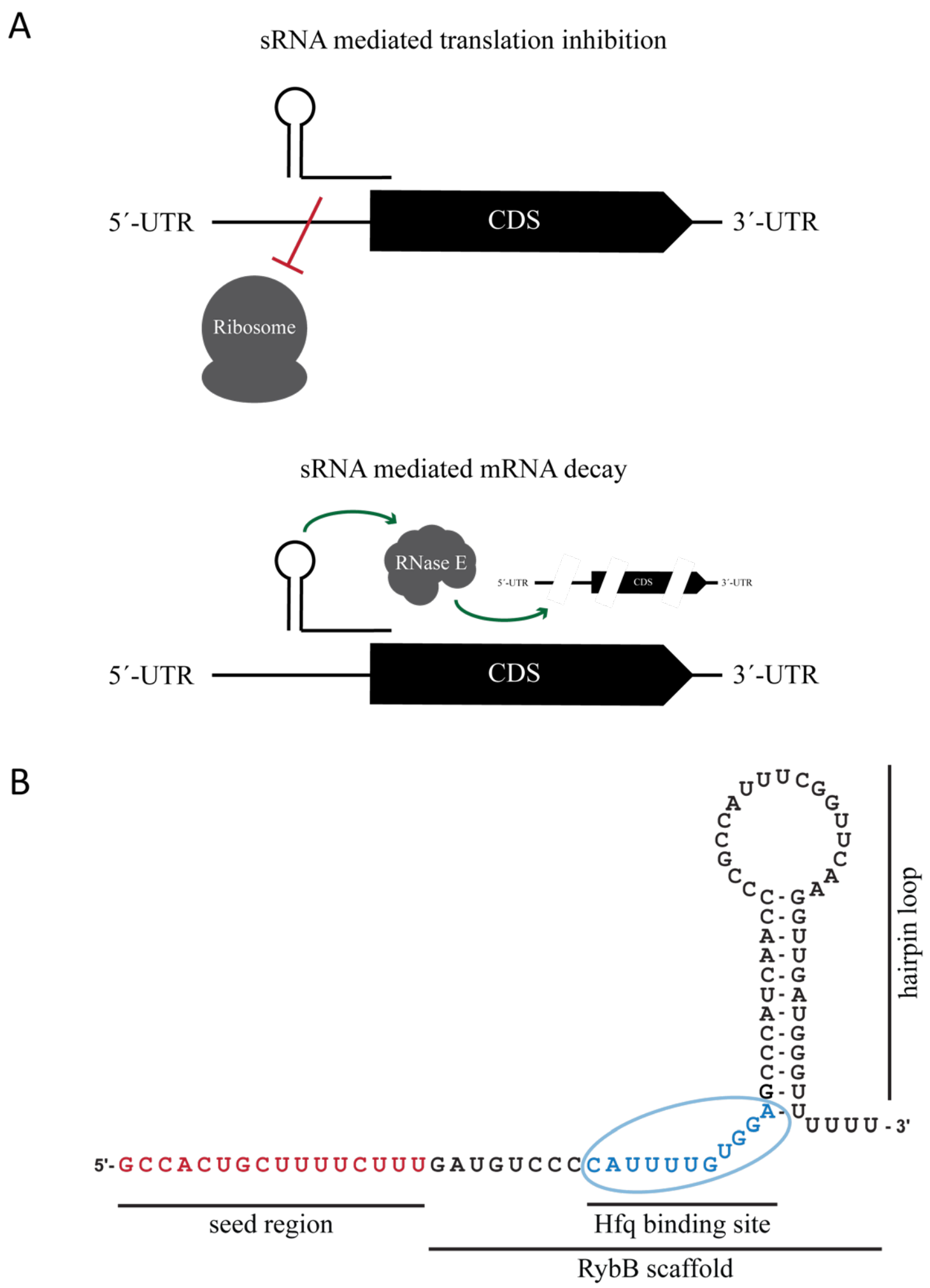

**Figure 1 | Function and modular structure of bacterial sRNAs.** (A) Inhibition of translation by an sRNA bound to its target mRNA (upper panel), and RNase E recruitment to facilitate mRNA decay (lower panel). UTR indicates untranslated region and CDS the coding sequence. For simplicity the TIR is not indicated. (B) RybB, a well characterized sRNA, shows three distinct features: a seed region for base-pairing with its target, a hairpin loop for transcription termination, and a Hfq binding site. The transcriptional terminator and Hfq binding site are part of the RybB scaffold.

Here, we provide (i) a detailed protocol for *in silico* prediction of seed regions for synthetic sRNA design, (ii) a modular cloning approach for the *in vitro* construction of sRNAs, and (iii) assays for the functional characterization of the synthetic sRNAs *in vivo* (Fig. 2A). Once the optimal sRNA is identified and characterized it can be used for its target application. The seed prediction tool, SEEDling (https://github.com/DIGGER-Bac/SEEDling), is used to predict optimal seed regions for a chosen target [10]. SEEDling integrates multiple parameters to reduce off-target effects and minimize the risk of structural changes in the synthetic sRNAs. The predicted seed regions are synthesized as oligonucleotides and used for *in vitro* construction of synthetic sRNA transcriptional units (TUs) by Golden Gate cloning. The chapter further provides strategies for the functional characterization of the constructed sRNA TUs based on phenotypic and/or fluorescent reporter assays [11]. In this protocol we are targeting the *acrA* mRNA, encoding a component of a major multidrug efflux pump in *E. coli* and relatives, as an example. The absence of AcrA increases susceptibility towards the $\beta$-lactam antibiotic oxacillin and allows for quick functional characterization of *acrA*-targeting sRNAs (Fig. 2B)[11]. The seed region of the synthetic sRNA ideally covers the TIR, consisting of the Shine-Dalgarno sequence and the start codon, to achieve efficient translational repression (Fig. 2C). Detailed information for *E. coli* endogenous transcriptional units can be found on https://ecocyc.org/ [12]. Notably, if repression of the targeted mRNA does not produce a phenotype, fluorescent reporters can be used as readout for sRNA functionality. The whole described workflow can be performed in any standard molecular biology laboratory.

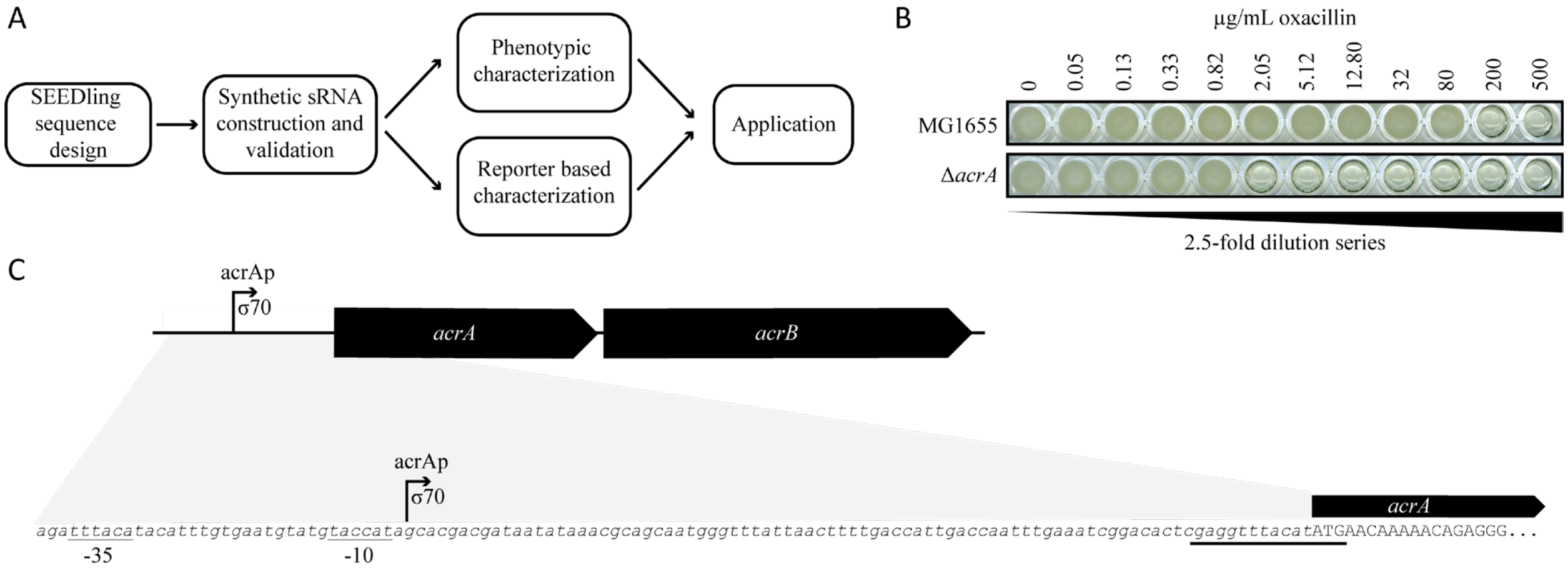

**Figure 2 | Workflow for synthetic sRNA construction, oxacillin susceptibility assay and structure of the *acrAB* operon.** (A) The general workflow of the presented protocol for *in silico* design, *in vitro* construction and *in vivo* application of synthetic sRNAs in bacteria. (B) Determination of the minimum inhibitory concentration (MIC) for oxacillin in *E. coli* wild type MG1655 and a **Δ***acrA* deletion strain. The increased oxacillin susceptibility of *E. coli* **Δ***acrA* can be used as a phenotypic readout for the functionality of synthetic sRNAs. The dilution series is performed in a 2.5-fold manner indicating an approximately 100-fold higher susceptibility of the **Δ***acrA* strain compared to the wild type. (C) Detailed visualization of the *acrAB* operon. The Shine-Dalgarno sequence and the first codons of the targeted mRNA are the ideal region to identify optimal seed regions for synthetic sRNAs. For visualization the TIR is underlined and the -35 and -10 regions are indicated.

## 2 Materials

The conduction of the described protocol requires the following standard laboratory equipment and consumables.

1. Standard reagents, consumables and instrumentation for PCR reactions
2. Standard reagents and equipment for gel electrophoresis
3. Standard reagents, consumables and instrumentation for microbial culturing
4. Standard reagents, consumables and instrumentation for transformation of *E. coli*
5. Standard reagents, consumables and instrumentation for plasmid extraction
6. Plate reader for continuous growth absorbance measurements
7. Plate reader for fluorescence and absorbance detection
8. Standard micropipettes and consumables, 12 channel pipettes are advised for work with microtiter plates

## 2.1 Computational equipment

Computer with Linux, macOS or Windows OS with Docker installed. The most recent SEEDling Docker image and its documentation is available at https://github.com/DIGGER-Bac/SEEDling under the CC BY-NC-SA 4.0 license. (*see* **Note 1**) R and the graphical interface RStudio are used for analysis (https://www.r-project.org/).

## 2.2 Plasmids

All plasmids relevant for this protocol are listed in Table 1. Plasmids can be requested from the corresponding author. Notably, any other Golden Gate cloning acceptor plasmids can be used. However, it is imperative to adjust the outlined protocol towards the cloning standard in regard to the desired type IIs enzymes, corresponding overhangs and plasmid features (e.g. antibiotic resistance).

**Tab. 1 | Plasmids constructed and used in this study.**

| Name | Relevant features | Parental plasmid | Reference |
|---|---|---|---|
| pSL009 | pBAD-TOPO derivative serving as empty plasmid control; $Kan^R$ | | [11] |
| pSL099 | Level 0 plasmid for subcloning of fragments to be released with SapI; $Spec^R$ | pMA60 [13] | this study |
| pSL123 | Level 0 plasmid containing RybB scaffold to be released with SapI; $Spec^R$ | pSL099 | this study |
| pSL135 | Level 0 plasmid containing $P_L$lacO-1 to be released with SapI; $Spec^R$ | pSL099 | this study |
| pSL137 | pBAD-TOPO derivative allowing Golden Gate cloning of TUs with SapI; $Kan^R$ | pSL009 | this study |
| pSL574 | pSL137 derivative containing RybB TU under $P_L$lacO-1 control with SEEDling seed sequence prediction S#1; $Kan^R$ | pSL137 | this study |
| pSL575 | pSL137 derivative containing RybB TU under $P_L$lacO-1 control with SEEDling seed sequence prediction S#2; $Kan^R$ | pSL137 | this study |
| pSL576 | pSL137 derivative containing RybB TU under $P_L$lacO-1 control with SEEDling seed sequence prediction S#3; $Kan^R$ | pSL137 | this study |
| pSL577 | pSL137 derivative containing RybB TU under $P_L$lacO-1 control with SEEDling seed sequence prediction S#4; $Kan^R$ | pSL137 | this study |
| pSL578 | pSL137 derivative containing RybB TU under $P_L$lacO-1 control with SEEDling seed sequence prediction S#5; $Kan^R$ | pSL137 | this study |
| pSL579 | pSL137 derivative containing RybB TU under $P_L$lacO-1 control with SEEDling seed sequence prediction S#6; $Kan^R$ | pSL137 | this study |
| pSL580 | pSL137 derivative containing RybB TU under $P_L$lacO-1 control with SEEDling seed sequence prediction S#7; $Kan^R$ | pSL137 | this study |
| pSL581 | pSL137 derivative containing RybB TU under $P_L$lacO-1 control with SEEDling seed sequence prediction S#8; $Kan^R$ | pSL137 | this study |

## 2.3 DNA oligonucleotides

Relevant oligonucleotides for the conduction of the protocol are provided in Table 2. 100 µM stocks are generated with $ddH_2O$ and stored at -20 °C. For PCR and sequencing reactions 10 µM working stocks are generated with $ddH_2O$ and stored at -20 °C.

**Tab. 2 | Oligonucleotides for the conduction of the described protocol.**

| Name | Sequence (5´-3´) * | Information |
|---|---|---|
| SLo3961 | CTGTCAAATGGACGAAGCAG | Colony PCR and sequencing primer |
| SLo3962 | CAGGCAAATTCTGTTTTATCAGACC | Colony PCR and sequencing primer |
| SLo5230 | **ATG**TTCATATGTAAACCTC | Forward primer of SEEDling predicted seed region S#1 targeting AcrA mRNA with corresponding Golden Gate cloning overhang indicated in bold. |
| SLo5231 | **ATG**GTTCATATGTAAACCT | Forward primer of SEEDling predicted seed region S#2 targeting AcrA mRNA with corresponding Golden Gate cloning overhang indicated in bold. |
| SLo5232 | **ATG**GCCAGAGGCGTAAACC | Forward primer of SEEDling predicted seed region S#3 targeting AcrA mRNA with corresponding Golden Gate cloning overhang indicated in bold. |
| SLo5233 | **ATG**TCATATGTAAACCTCG | Forward primer of SEEDling predicted seed region S#4 targeting AcrA mRNA with corresponding Golden Gate cloning overhang indicated in bold. |
| SLo5234 | **ATG**CCCTCTGTTTTTGTTC | Forward primer of SEEDling predicted seed region S#5 targeting AcrA mRNA with corresponding Golden Gate cloning overhang indicated in bold. |
| SLo5235 | **ATG**CGCCAGAGGCGTAAAC | Forward primer of SEEDling predicted seed region S#6 targeting AcrA mRNA with corresponding Golden Gate cloning overhang indicated in bold. |
| SLo5236 | **ATG**TGTTCATATGTAAACC | Forward primer of SEEDling predicted seed region S#7 targeting AcrA mRNA with corresponding Golden Gate cloning overhang indicated in bold. |
| SLo5237 | **ATG**GACCGCCAGAGGCGTA | Forward primer of SEEDling predicted seed region S#8 targeting AcrA mRNA with corresponding Golden Gate cloning overhang indicated in bold. |
| SLo5238 | **ATC**GAGGTTTACATATGAA | Reverse primer of SEEDling predicted seed region S#1 targeting AcrA mRNA with corresponding Golden Gate cloning overhang indicated in bold. |
| SLo5239 | **ATC**AGGTTTACATATGAAC | Reverse primer of SEEDling predicted seed region S#2 targeting AcrA mRNA with corresponding Golden Gate cloning overhang indicated in bold. |
| SLo5240 | **ATC**GGTTTACGCCTCTGGC | Reverse primer of SEEDling predicted seed region S#3 targeting AcrA mRNA with corresponding Golden Gate cloning overhang indicated in bold. |
| SLo5241 | **ATC**CGAGGTTTACATATGA | Reverse primer of SEEDling predicted seed region S#4 targeting AcrA mRNA with corresponding Golden Gate cloning overhang indicated in bold. |
| SLo5242 | **ATC**GAACAAAAACAGAGGG | Reverse primer of SEEDling predicted seed region S#5 targeting AcrA mRNA with corresponding Golden Gate cloning overhang indicated in bold. |
| SLo5243 | **ATC**GTTTACGCCTCTGGCG | Reverse primer of SEEDling predicted seed region S#6 targeting AcrA mRNA with corresponding Golden Gate cloning overhang indicated in bold. |
| SLo5244 | **ATC**GGTTTACATATGAACA | Reverse primer of SEEDling predicted seed region S#7 targeting AcrA mRNA with corresponding Golden Gate cloning overhang indicated in bold. |
| SLo5245 | **ATC**TACGCCTCTGGCGGTC | Reverse primer of SEEDling predicted seed region S#8 targeting AcrA mRNA with corresponding Golden Gate cloning overhang indicated in bold. |

* bold letters indicate the overlaps for Golden Gate cloning

## 2.4 Enzymes

Any enzyme with corresponding properties can be used. For the presented protocol all used enzymes were provided by New England Biolabs (NEB). (*see* **Note 2**)

1. T4 DNA Ligase (400,000 units/mL)
2. SapI (10,000 units/mL)
3. Taq DNA Polymerase (5,000 units/mL)

## 2.5 Antibiotics

All antibiotics used in this study are solved in sterile $H_2O$ and are stored in 1 mL aliquots at -20 °C. The stock concentration is 1,000 x except for oxacillin which was used in the indicated concentrations.

1. Kanamycin (50 mg/mL stock)
2. Oxacillin (50 mg/mL stock)
3. Spectinomycin (120 mg/mL stock)

## 2.6 Chemicals, buffers and media components

1. LB medium (1% (w/v) tryptone, 0.5% (w/v) yeast extract, 1% (w/v) sodium chloride, pH 7.0 ± 0.2)
2. 10 x annealing buffer (10 mM Tris(-HCl), pH 7.5-8.0, 50 mM NaCl and 1 mM EDTA)
3. 1 x TAE buffer (1 mM EDTA · $Na_2$ · 2 $H_2O$, 20 mM acetate, 40 mM Tris)
4. DNA dye: Thiazole orange dissolved in dimethyl sulfoxide (DMSO) (10,000 x stock concentration: 13 mg/mL)
5. Agarose (standard)
6. DNA ladder
7. Solid media is prepared with 2% Agar

## 2.7 Consumables

1. Single well microtiter plates (e.g. PlusPlates (PLU-003), Singer Instruments)
2. 96-well microtiter plates
3. Optical clear microtiter plate seal
4. PCR reaction tubes
5. Standard petri dishes

## 2.8 Strains

All relevant laboratory *E. coli* strains for the outlined protocol are provided in Table 3.

**Tab. 3 | Strains used in this study.**

| Name | Relevant features | Reference |
|---|---|---|
| *E. coli* MG1655 | K-12 F− λ− | [14] |
| Δ*acrA* | *E. coli* MG1655 Δ*acrA::cat*; $Cm^R$ | [11] |
| *acrA-9'−syfp2* | *E. coli* MG1655 *acrA-9'-syfp2-cat*, translational fusion of first 9 *acrA* codons; $Cm^R$ | [11] |

# 3 Methods

## *3.1 Seed prediction using SEEDling*

This part of the methods section explains the use of SEEDling (https://github.com/DIGGER-Bac/SEEDling) [10] for the prediction of seed regions for mRNA targeting. The process is performed by targeting the *acrA* mRNA of *E. coli*. However, SEEDling can be adjusted to any target organism and target mRNA. SEEDling is capable of generating global predictions for whole genomes. The predicted sequences will be used for the downstream steps of the protocol. The workflow for seed prediction using SEEDling is visualized in Figure 3A.

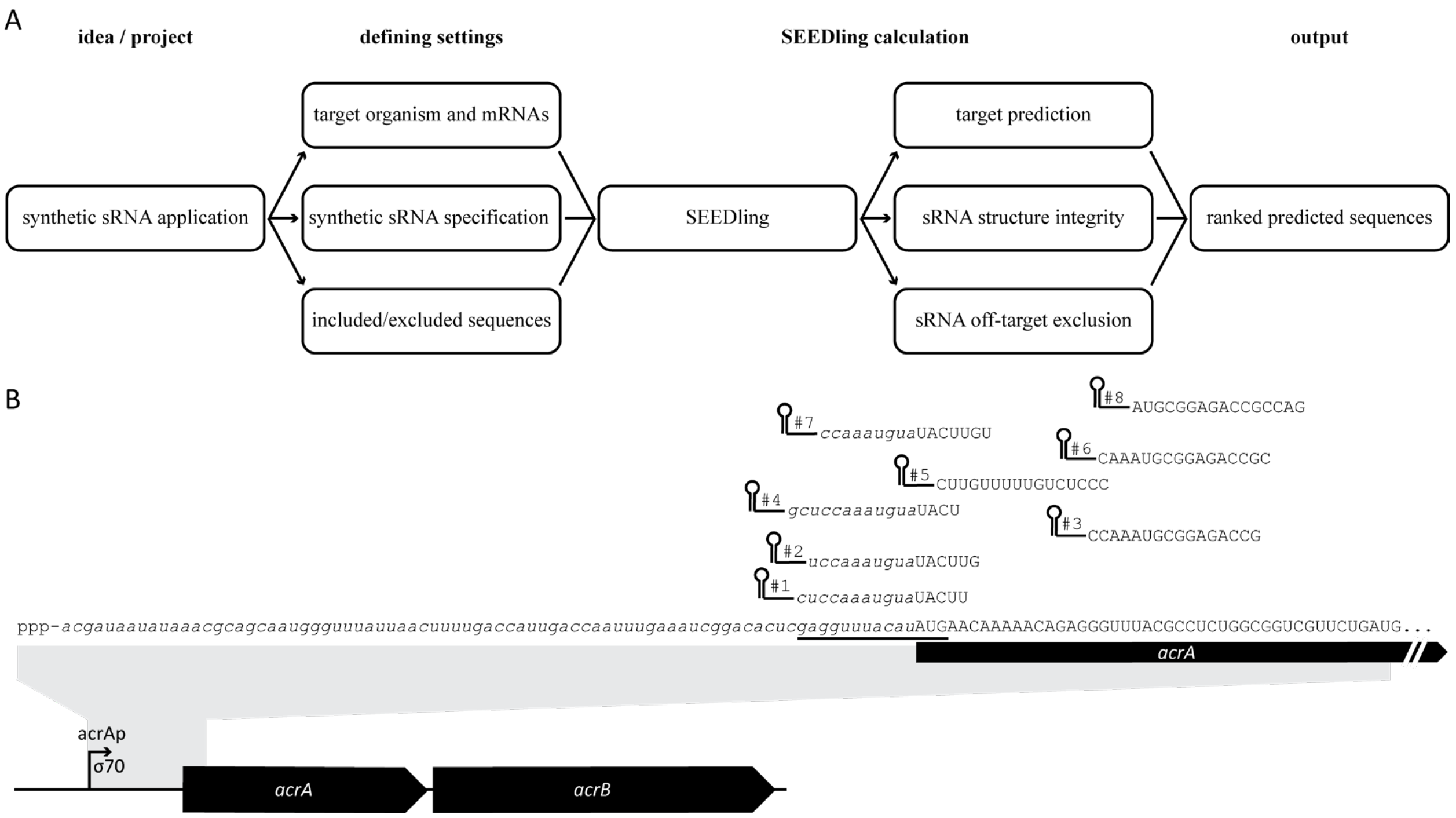

**Figure 3 | Prediction of seed regions using SEEDling.** (A) The workflow of SEEDling is divided into four steps. The initial idea and project conceptualization is followed by setting sRNA criteria by the user. The settings are used by SEEDling to predict the best possible seed regions by keeping the sRNA structure integrity and prevent potential off-target effects. SEEDling provides an output file with detailed information in regard to each prediction to allow the user the best possible selection. (B) Visualization of synthetic sRNAs binding to *acrA* mRNA in *E. coli*. Seed regions with a length of 16 nucleotides are based on SEEDling predictions. The number for each synthetic sRNA indicates the ranking within the SEEDling prediction. The 5´-UTR is visualized in italics, the coding sequence in capital letters and the TIR is underlined.

1. Install the most recent version of SEEDling based on the documentation on GitHub (https://github.com/DIGGER-Bac/SEEDling). The use of the Docker image is the preferred use of SEEDling to prevent incompatibility issues. Prior to the use of the SEEDling Docker image, the compatible Docker Desktop App must be installed (https://www.docker.com/products/docker-desktop/). (*see* **Note 3**)

2. For Seedling prediction, the following files in the input folder need to be adjusted towards the prediction goal:

    1. config.yml
    2. exclude.fasta
    3. include.txt
    4. reference_file.gb

    (a) The *configuration* file (config.yml) is necessary to be adjusted towards the prediction goals (e.g., target gene, reference genome, seed region length, intended scaffold). The here described protocol generates predictions to target the *acrA* mRNA (Fig. 3B). (b) The *exclude* file contains a list of DNA sequences in fasta format to be excluded from the prediction (e.g., type IIs recognition sites). (c) The *include* file contains the list of target genes selected from the provided

reference genome. (d) The *reference* file contains the annotated genome of the target organism in GenBank format (*see* **Note 4**).

All files can be edited with a standard text file editor (e.g., Editor in Windows OS or Notepad++, https://notepad-plus-plus.org/).

3. Adjust the config.yml according to the desired output. The individual settings are explained in detail in the SEEDling GitHub documentation. For this protocol the following settings were used:

```
# Path to GenBank target genome (used in off-target detection)
subject_path: "input/e_coli_k12_mg1655.gb"
# Path to GenBank files of genes for which the SEED will be calculated
target_path: "input/e_coli_k12_mg1655.gb"
# Path to output file, .csv
output_path: "input/test.csv"
# Number of SEED predictions per gene
select_top: 8
# Offset (left of the start codon)
start_offset: 40
# Offset (right of the start codon)
end_offset: 20
# Step size (for sliding window)
step_size: 1
# Length of the SEED region
seq_length: 16
# Prefix (Scaffold)
srna_prefix: ""
# Suffix (Scaffold)
srna_suffix: "GATGTCCCCATTTTGTGGAGCCCATCAACCCCGCCATTTCGGTTCAAGGTTGATGGGTTTTTTGTT"
# Scaffold+SEED which should be used as a reference for RNApdist
srna_template:
"GCCACTGCTTTTCTTTGATGTCCCCATTTTGTGGAGCCCATCAACCCCGCCATTTCGGTTCAAGGTTGATGGGTTTTTTGTT"
# FASTA file of sites that should be excluded in the final scaffold+SEED
exclude_sequences_path: "input/exclude.fasta"
# Newline separated txt file of gene names which should be included (whitelist)
include_genes_path: "input/include.txt"
# E-value for BLASTn off target checks
blast_evalue: 0.04
```

4. Adjust the exclude.fasta according to the desired output. (*see* **Note 5**) For this protocol the following settings were used:

```
>SapI
GCTCTTC
```

5. Adjust the include.fasta according to the desired output. (*see* **Note 6**) For this protocol the following settings were used:

```
acrA
```

6. Provide the reference of your target organism in the form of a GenBank file. For this protocol the NC_000913 reference of *E. coli* was used to generate a GenBank file ("e_coli_k12_mg1655.gb"). Besides providing the target gene sequences for seed predictions, the reference is further used for off-target predictions.

7. Open the Docker Desktop application. Load the SEEDling Docker image via terminal, cmd or powershell with the command: (*see* **Note 7**)

```
docker load --input seedling
```

8. Start SEEDling Docker image to open a Docker container by choosing the ‘run’ option in the Docker Images tab.

   ***Important:*** When creating the container set the *container path* in the optional settings to "/home/DIGGER/SEEDling/input". The *host path* can be freely adjusted by the user.

9. In the Docker Containers tab: open the terminal of the SEEDling container and execute the prediction program which generates the output as a csv file. Command:

```
python3 SEEDling.py -c input/config.yml
```

10. The output file is printed as test.csv into the input folder (as defined in the config.yml file). The output file is used to identify ideal seed sequences to be subsequently ordered as oligonucleotides. (*see* **Note 8**) For the settings used in this protocol, eight seed sequences were predicted and ordered as forward and reverse oligonucleotides for subsequent annealing to generate double stranded DNA for Golden Gate cloning. (*see* **Note 9**) Figure 3B visualizes the position of the individual predicted seed sequences and Table 4 provides output information for seed region S#1.

**Tab. 4 | SEEDling output data for predicted seed region S#1.**

| | |
|---|---|
| ID | ZLZRJ1XP |
| Source | input/e_coli_k12_mg1655.gb |
| RNAdist | 3.86201 |
| Hybrid Energy | -34.48 |
| Prefix | - |
| Suffix | GATGTCCCCATTTTGTGGAGCCCATCAACCCCGCCATTTCGGTTCAAGGTTGATGGGTTTTTTGTT |
| Seed | TTCATATGTAAACCTC |
| Gene | *acrA* |
| Start | 485614 |
| End | 485630 |
| Strand | -1 |
| Offsite | FALSE |
| Illegal Site | FALSE |
| Fold | .......................((((...))))((((((((((((..(((.....)))....)))))))))))))....... |
| FullSeq | TTCATATGTAAACCTCGATGTCCCCATTTTGTGGAGCCCATCAACCCCGCCATTTCGGTTCAAGGTTGATGGGTTTTTTGTT |

### *3.2 Construction and validation of synthetic sRNA TUs by Golden Gate cloning*

The predicted sequences of SEEDling are ordered as oligonucleotides and used in this protocol to generate synthetic sRNA TUs in a pBAD derivative (pSL137) by Golden Gate cloning in combination with previously generated parts for the $P_L$lacO-1 promoter (pSL135) and RybB scaffold (pSL123). (*see* **Note 10**) This section describes the Golden Gate cloning procedure in a stepwise manner. The protocol assumes that the individual parts in level 0 plasmids and the acceptor plasmid are available to the user. However, this procedure can be adapted to any other Golden Gate cloning system. The only requirements are the adjustment of the type IIs enzyme and the corresponding overhangs for the sRNA TU cloning. Golden Gate cloning, the acceptor plasmid, Golden Gate reaction and the workflow is depicted in Figure 4.

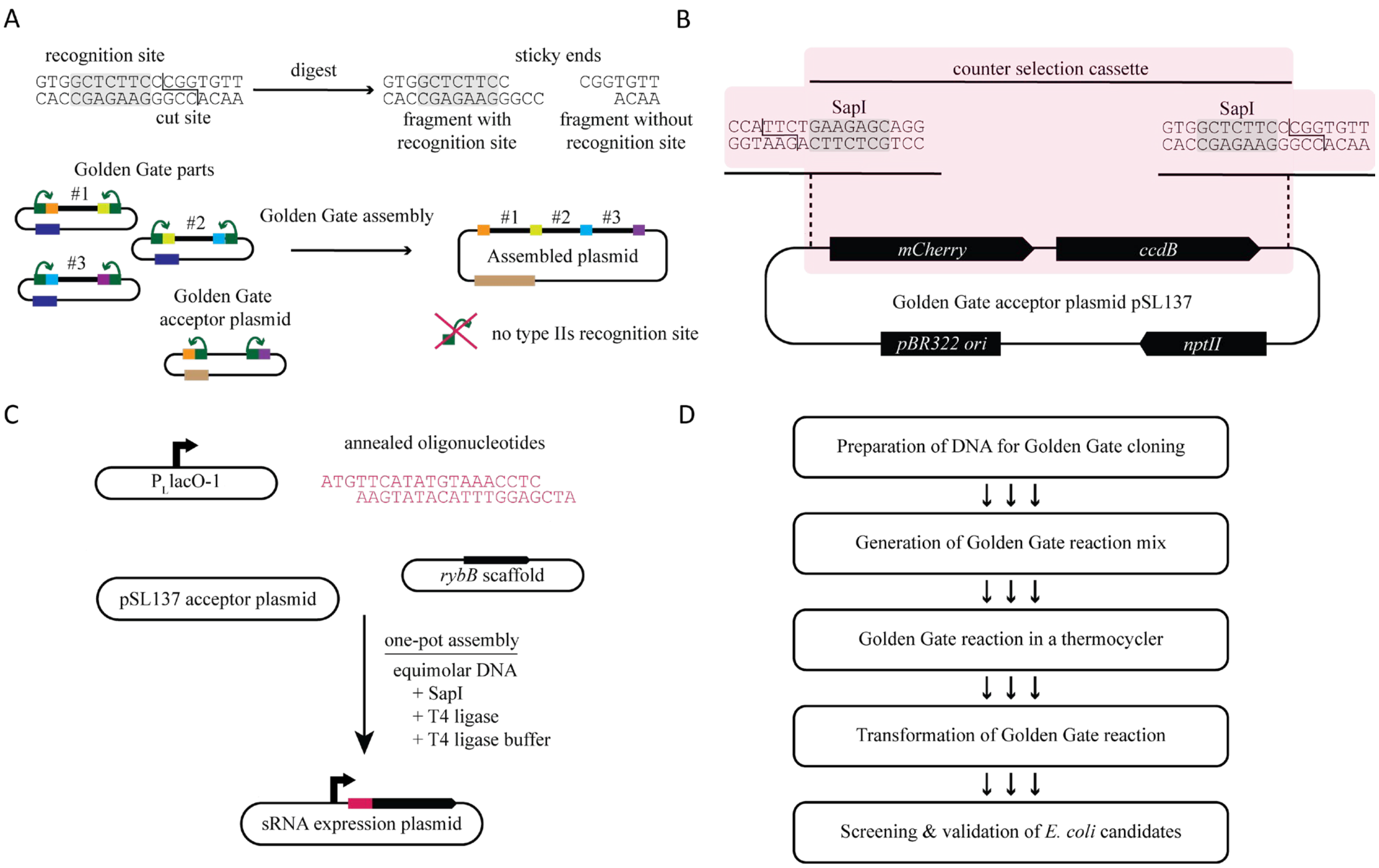

**Figure 4 | Golden Gate cloning of synthetic small RNAs.** (A) Visualization of Golden Gate cloning principle. The cloning strategy relies on the use of type IIs restriction enzymes which have a determined recognition site but cut an undefined DNA sequence in a fixed distance (upper panel). If type IIs recognition sites and resulting overhangs are orchestrated (indicated by colors), they can be used to perform a multi-fragment assembly into a corresponding acceptor plasmid (lower panel). If the assembly is planned properly, the resulting plasmid has the fragments in the right order and no more recognition sites of the used enzyme are present (green box with arrows). Level 0 parts are released from the plasmid with type IIs recognition sites in the plasmid backbone while the acceptor plasmid has the type IIs recognition sites in the opposite orientation. It is imperative that level 0 plasmids and the acceptor plasmid have different antibiotic selection markers indicated by the dark blue and beige bar respectively. (B) Easy-to-use acceptor plasmid based on Koebel *et al*. 2022 containing a dual-selection cloning cassette in a pBAD derivative. pSL137 differs from previous plasmids by relying on the use of the SapI type IIs enzyme instead of BbsI. SapI generates 3-nucleotide overhangs instead of 4-nucleotide overhangs (BbsI). (C) Golden Gate cloning strategy used within this protocol. Two basic part plasmids ($P_L$lacO-1 promoter [pSL123] and *rybB* scaffold [pSL135]) and the respective annealed oligonucleotides (red) are assembled with the acceptor plasmid pSL137 in a one-pot Golden Gate reaction using SapI, resulting in the respective sRNA expression plasmid. The arrow depicts the promoter and the red bar in the assembled plasmid depicts the seed region of the synthetic sRNA TU. (D) The workflow of this protocol section includes the preparation and conduction of the Golden Gate reaction with subsequent transformation into *E. coli* cells and plasmid validation. The stepwise procedure is given in the text. If all parts are available, the reaction and transformation into *E. coli* can be achieved within one day. Single colonies can be isolated on the second day and plasmids for validation can be extracted on the third day. This pipeline allows characterization and application of synthetic sRNA TUs starting on the fourth day.

### *3.2.1 Golden Gate cloning*

1. Generate double-stranded DNA from single-stranded oligonucleotides by annealing.

2. The two corresponding oligonucleotides are annealed by mixing 4.5 µL of each 100 µM oligonucleotide with 1 µL of 10 x annealing buffer in a PCR microcentrifuge tube. Annealing is performed in a thermocycler by heating the respective sample(s) to 95 °C for 5 minutes and cooling it to 25 °C with the lowest possible ramp speed. Alternatively, the reaction can be performed by boiling the mixture(s) in a heat block or water bath and slowly cooling down to room temperature at the bench.

3. Add 100 µL dd$H_2O$ to the annealed oligos. (*see* **Note 11**)

4. Measure and prepare other DNA (*see* **Note 12**) parts such as the acceptor plasmid (pSL137) and level 0 plasmids (pSL123 and pSL135). (*see* **Note 13**)

5. Mix the following components in a PCR microcentrifuge tube for the Golden Gate reaction: (*see* **Note 14**)

| **Compound** | **Volume** |
|---|---|
| T4 DNA Ligase Buffer | 1 µL |
| T4 DNA Ligase (400,000 units/mL) | 1 µL |
| Type IIs restriction enzyme (here: SapI) | 1 µL |
| Acceptor plasmid | ~20 fmol (*see* **Note 15**) |
| Level 0 plasmid | each ~20 fmol |
| Annealed oligonucleotides | 20 fmol up to 2 pmol |
| dd$H_2O$ | ad 10 µL |

6. Mix the reaction briefly by flipping the tube or pulse vortexing. Spin reactions briefly in a microcentrifuge ensuring the reaction mix is at the bottom of the tube and place the reaction in a thermocycler running the following program:

| **Temperature [°C]** | **Time [min]** |
|---|---|
| 37 | 300 |
| 50 | 20 |
| 80 | 10 |
| 8 | ∞ (storage) |

7. Use the Golden Gate reaction directly for transformation into the *E. coli* host strain or store reaction mix at -20 °C for subsequent transformation. (N)

### *3.2.2 Transformation of Golden Gate reaction into E. coli cells*

Within this protocol we use the transformation into in-house generated competent cells. Cells are generated according to a RbCl based method [15]. However, other methods are suitable as well and can be adjusted to the preferences of the user. In our case, *E. coli* wild type MG1655 cells are used allowing direct testing of synthetic sRNA TUs.

1. Transform chemically competent *E. coli* cells with Golden Gate reaction mix. For this purpose, add 2-10 µL reaction mix to 20 µL competent cells in a 1.5 mL microcentrifuge tube. Mix carefully by flipping the microcentrifuge tube.

2. Incubate mixture on ice for 30 minutes. Heat up a water bath to 42 °C.

3. Heat shock: Place the mixture in the preheated water bath for 30 seconds.

4. Put the reaction mix back on ice for 5 minutes.

5. Add 1 mL of LB medium (ideally preheated to 37 °C).

6. Incubate at 37 °C with a shaking frequency of 180 rpm for 60 minutes.

7. Centrifuge the tube at 7,000 x g for 3 minutes to obtain a cell pellet.

8. Discard part of the supernatant and retain a volume of approximately 100 µL.

9. Resuspend the cell pellet by gently pipetting up and down.

10. Plate the cell suspension on LB agar with the corresponding antibiotic for plasmid maintenance using sterile glass beads or a cell spreader.

11. Incubate plates overnight at 37 °C. (*see* **Note 16**)

### *3.2.3 Screening and validation of plasmid DNA*

In this protocol colony PCR (cPCR) is used to identify potentially correct candidates. Subsequently extracted plasmid DNA is sent for external Sanger sequencing services to validate the integrity of the DNA sequence.

1. Add 50 µL of sterile ddH$_2$O to a PCR microcentrifuge tube.

2. With a sterile toothpick or pipette tip, pick a colony of the transformation plate and put it in the prepared tube. Stir the toothpick or pipette tip to suspend cells in the water. Store at 4 °C after use. (*see* **Note 17**)

3. On ice, prepare cPCR reaction master mix according to the number of screened candidates. Plan at least 10% dead volume for the master mix preparation. Dispense 9 µl of master mix into PCR reaction tubes or the wells of a 96-well PCR plate. Add 1 µl of candidate cell suspension to the corresponding wells. Example components of the master mix for a single reaction:

| **Component** | **1-fold Volume** |
|---|---|
| Forward cPCR primer (10 µM) | 0.2 µL |
| Reverse cPCR primer (10 µM) | 0.2 µL |
| 2 x Polymerase Master Mix including dNTPs and colored buffer for direct gel loading | 5 µL |
| ddH$_2$O | 3.6 µL |

4. In a thermocycler, use the following settings:

| **Description** | **Cycles** | **Temperature** | **Time** |
|---|---|---|---|
| Initial denaturation | 1 | 94 °C | 5 minutes |
| Denaturation<br>Annealing<br>Elongation | 35 | 94 °C<br>45-68 °C<br>68 °C | 15-30 seconds<br>15-60 seconds<br>60 seconds/kb |
| Final extension | 1 | 68 °C | 5 minutes |
| Storage | 1 | 8 °C | ∞ |

5. Prepare a 1-2% (w/v) agarose gel depending on expected fragment size. Mix the agarose with a DNA dye (e.g., thiazole orange (*see* **Note 18**)).

6. Perform agarose gel electrophoresis with the whole cPCR reaction mix and a respective DNA ladder in 1 x TAE buffer at 100 V for ~40 minutes. (*see* **Note 19**)

7. Visualize and document results of the gel electrophoresis with a dedicated documentation system.

8. Inoculate potentially correct candidates containing desired cloning products in 5 mL LB with the respective antibiotic for plasmid maintenance and incubate overnight at 37 °C with shaking at 180 rpm.

9. Purify plasmids of candidates with method of choice. (*see* **Note 20**)

10. Verify integrity of plasmid sequences by Sanger sequencing using an appropriate sequencing primer.

11. ***Optional:*** Preserve candidates as cryo-culture by combining 700 µL of dense early stationary phase culture with 300 µL 50% glycerol and store at -70 °C in a cryo-tube.

### *3.3 Synthetic sRNA functionality test on solid media*

Repression of *acrA* results in oxacillin susceptibility of *E. coli* cells allowing for a functional characterization of the constructed synthetic sRNA TUs. The initial test is performed on solid media with increasing oxacillin concentrations to obtain a qualitative assessment of synthetic sRNA functionality. Figure 5 gives an overview of the workflow and shows exemplary results.

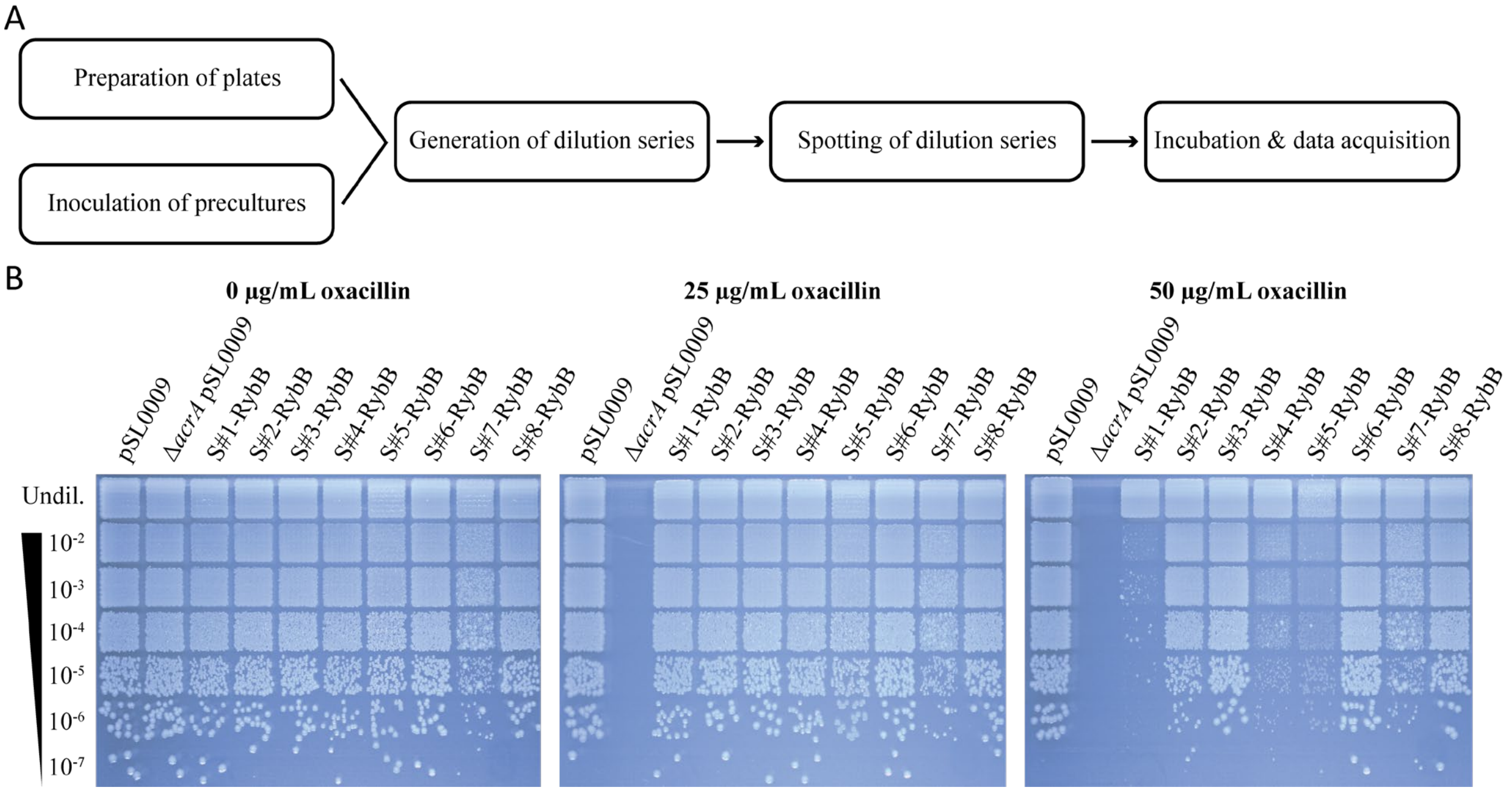

**Figure 5 | Phenotypic testing of the constructed synthetic small RNAs on solid agar.** (A) The workflow is depicted with the detailed steps given in the protocol. The procedure consists of five main steps which are the preparation of cultures and plates, generation and spotting of dilution series, and data acquisition. (B) Oxacillin susceptibility assay on solid medium was performed using the Singer Instruments Rotor HDA+ and the 7x7 spotting program. A 96-well microtiter plate prepared with the indicated dilution series was prepared in sterile distilled $H_2O$ and spotted on LB agar in single-well plates containing kanamycin for plasmid maintenance and the indicated oxacillin concentration. The Δ*acrA* strain shows high susceptibility to oxacillin. The synthetic sRNAs S#1-, S#4- and S#5-RybB show reduced growth at 50 µg/mL oxacillin. For sRNAs S#2-, S#3-, S#6- and S#8-RybB no regulation can be observed. Synthetic sRNA S#7-RybB indicates off-target effects by a small colony phenotype already in the absence of oxacillin, which underlines the importance to benchmark synthetic sRNAs prior to their application. The results

are consistent with the alternative benchmarking procedures in form of liquid growth and fluorescence reporter assays indicating that one type of benchmarking may be sufficient for sRNA characterization.

1. If plasmids are not in the desired strain background, transform into respective *E. coli* cells and make sure respective controls are generated (here: *E. coli* wild type with pSL009 and *E. coli* Δ*acrA* with pSL009).

2. Prepare single-well microtiter plates with agar. Use a 50 mL conical tube, fill it with 40 mL molten LB-agar, containing the appropriate antibiotic for plasmid maintenance, and add the respective volume of compound for the phenotypic screening (here: oxacillin at 0 to 100 µg/mL)( *see* **Note 21**). Always include a control without the compound.

   ***Important:*** The conical tube can be reused if compound concentration increases.
   ***Important:*** Handle the warm molten medium with care.

3. Close the conical tube and invert it carefully 2-3 times without creating bubbles to equally distribute the antibiotic(s) in the medium.

4. Carefully pour medium into a single-well microtiter plate and make sure that plates are not moved until properly solidified to prevent uneven surfaces. (*see* **Note 22**)

5. After plates are solidified, store upside down in an airflow free area at room temperature overnight. Protect plates from direct sunlight. If plates are not used the following day, store in a plastic bag at 4 °C.

6. Grow precultures of dedicated strains from single colonies overnight in LB medium containing the appropriate antibiotic to maintain the sRNA expression plasmid. Always include dedicated positive and negative controls (here: *E. coli* wild type with pSL009 and *E. coli* Δ*acrA* with pSL009)

7. Transfer 200 µL of the dedicated strains in row A of a 96-well microtiter plate.

8. Dispense 198 µL sterile $H_2O$ into row B and 180 µL in the remaining rows of the 96-well microtiter plate.

9. Dilute the first row 1:100 into the second row. From the second row onwards perform 1:10 dilutions. (*see* **Note 23**)

10. In this protocol the Singer Instruments Rotor HDA+ was used to perform spotting onto solid agar plates utilizing the 7x7 spotting program creating a grit of 49 spots for 96 samples in parallel (Fig. 5B). If no Rotor HDA+ is available, spotting can be performed using multichannel pipettes. Carefully transfer up to 5 µl of the corresponding dilution onto the agar plate. (*see* **Note 24**)

11. Incubate the spotted plates at 37 °C and document growth by imaging at dedicated time points (e.g. overnight and 24 hours). In this protocol the Singer Instruments PhenoBooth was used. Any other photo documentation system is suitable as well.

12. Perform qualitative evaluation of acquired data. (*see* **Note 25**)

### 3.4 Synthetic sRNA functionality test in liquid media

The parallel growth in a plate reader of many synthetic sRNA expressing *E. coli* strains can be utilized for functional characterization. The corresponding workflow is depicted in Figure 6A. The comparison of the area under the curve (AUC) of *E. coli* cultures containing a respective compound and cultures containing no compound allows for a more quantitative characterization of the synthetic sRNAs in comparison to the respective controls (Fig 6B). The AUC was found to be more precise in comparison to maximum cell density and generation time because potential suppressor mutants may overgrow the culture or cell filamentation may be induced by the compound [11]. Results for the generated synthetic sRNAs within this protocol are shown in Figure 6C.

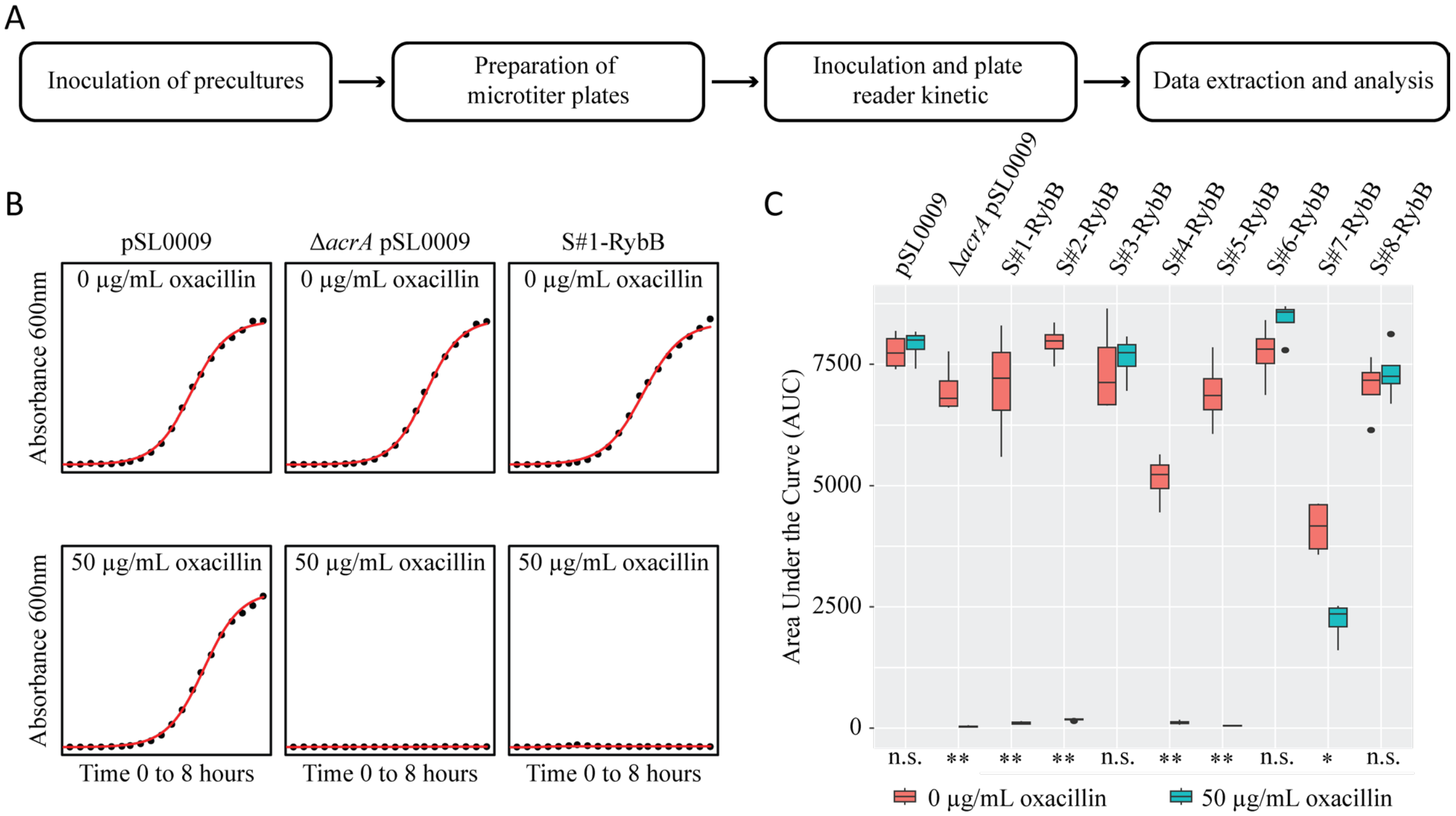

**Figure 6 | Phenotypic testing of the constructed synthetic small RNAs in liquid media.** (A) The workflow is depicted with the detailed steps given in the protocol. The procedure consists of four main steps which are the preparation of cultures and plates, data acquisition and data analysis. (B) Exemplary 'growthcurver' output showing microbial growth for a single well in the presence and absence of 50 µg/mL oxacillin. Dots represent individual measurements and the red line represents the fitted curve. Based on the fitted curve 'growthcurver' determines the area under the curve (AUC). The S#1-RybB growth curve in the presence of oxacillin shows a small peak which becomes more apparent at lower oxacillin concentrations (data not shown). This effect is caused by a filamentation phenotype of bacterial cells. (C) Visualization of oxacillin liquid growth test. AUC results for the synthetic sRNA expression strains are shown in comparison to the wild type and the Δ*acrA* strain. Measurements are performed in quadruplicates for each condition. Red boxes show AUC in the absence of oxacillin and turquoise boxes AUC in the presence of 50 µg/mL oxacillin. Student's t-test was applied for statistical evaluation (n.s.: not significant, *: $P < 0.01$, **: $P < 0.001$). Synthetic sRNAs S#1-, S#2-, S#4- and S#5-RybB cause a strong growth defect in the presence of oxacillin. Synthetic sRNAs S#3-, S#6- and S#8-RybB have almost no effect on growth. Targeting the TIR of *acrA* - instead of the coding region - seems to provide optimal sRNA functionality (*cf* Fig. 3B). Synthetic sRNA S#7-RybB already shows a growth defect in the absence of oxacillin indicating off-target effects of this sRNA. The liquid media results are consistent with the alternative benchmarking procedures except for S#2-RybB which does not indicate regulation on solid media. However, the different assays indicating that one type of benchmarking may be sufficient for sRNA characterization.

1. Inoculate candidates and controls from single colonies and grow overnight in LB medium containing the appropriate antibiotic for plasmid maintenance. (*see* **Note 26**)

2. Prepare a 96-well microtiter plate for optical measurements with 150 µL of LB medium containing the antibiotic for plasmid maintenance. (*see* **Note 27**)

3. If preculture was already prepared in a 96-well microtiter plate, it is recommended to use a 96-pin replica-plating stamp or if available a Singer Instruments Rotor HDA+ for inoculation of the experiment.

4. Seal plate with an optical clear adhesive seal or if available a thermal plate sealer with optical clear seal and the respective settings.

   ***Important:*** Be careful with hot surfaces if heat sealing is used.
   ***Important:*** Handle sealed plates with care to avoid splashes on the seal.

5. Use plate reader with kinetic settings to record microbial growth for at least 12 hours at 37 °C. Adjust settings according to the experimental setup if necessary. (*see* **Note 28**)

6. The data are exported as a csv file for further analysis after finishing data acquisition.

7. Analyze the data with your preferred analysis pipeline. Here, we use the open source statistic software R (https://www.r-project.org/) with the graphical user interface RStudio to analyze and visualize the generated data.

8. Install and load package 'growthcurver' [16].

```
install.packages('growthcurver')
library('growthcurver')
```

9. Prepare the input data as shown in the 'growthcurver' documentation (https://cran.r-project.org/web/packages/growthcurver/vignettes/Growthcurver-vignette.html).

10. Exemplary data format for a 96-well plate saved as a csv file containing the column headers in the first row:

| time | A1 | B1 | C1 | D1 | E1 | F1 | G1 | H1 | A2 | […] | H12 |
|---|---|---|---|---|---|---|---|---|---|---|---|
| x | 0.05 | 0.05 | 0.10 | 0.09 | 0.06 | 0.08 | 0.05 | 0.05 | 0.06 | […] | 0.05 |

11. The 'growthcurver' application generates a graphical output and corresponding values including the AUC which is used for the functional characterization of synthetic sRNAs.

12. Following commands can be used for the output generation:

```
# Load data in form of csv file
data <- read.csv(file.choose(), header = TRUE, sep = ",", stringsAsFactors = FALSE)
# Generate visual output as pdf file
gc_out <- SummarizeGrowthByPlate(data, plot_fit = TRUE, plot_file = "data.pdf")
# Inspect output data (optional)
View(gc_out)
```

13. Following commands can be used to get access to the AUC values as a text file:

```
# Name the output file and add a directory to save the data
output_file_name <- "the/path/to/my/data/myfilename.txt"
# Create the output file
write.table(gc_out, file = output_file_name, quote = FALSE, sep =  "\t", row.names = 
FALSE)
```

14. The auc_l (area under the logistic curve) values are taken from the 'growthcurver' output file for further analysis.

15. To analyze the effect of synthetic small RNAs after the antibiotic treatment, generating boxplots for clear visualizations using the R-package 'ggplot2' is recommended.

16. Install and load package 'ggplot2'.

```
install.packages('ggplot2')
library('ggplot2')
```

17. Prepare the input data as a csv file.

18. Example:

| **Sample** | **Treatment** | **AUC** |
|---|---|---|
| Control | No oxacillin | 5000 |
| Control | Oxacillin | 5000 |
| Sample S#1 | No oxacillin | 5000 |
| Sample S#1 | Oxacillin | 100 |

19. Following commands can be used to obtain boxplots as output:

```
# Load data in form of csv file
data <- read.csv(file.choose())
# Generate output
ggplot(data = name, aes(x = Sample,y = AUC, fill = Treatment)) + geom_boxplot()
```

20. In RStudio, the output can be visualized in the "Plots" tab and exported as a pdf or image.

21. Evaluate significance of results with a two-sided Student's t-test.

### *3.5 Fluorescence reporter assay*

The post-transcriptional regulation of a target mRNA may not necessarily produce a phenotype. To validate synthetic sRNA functionality, fluorescence reporter constructs can be utilized for assessment of translational repression. This part of the protocol outlines the respective steps based on an *acrA-9'-syfp2* translational fusion where s*yfp2* replaces the *acrA* coding sequence after the 9$^{th}$ codon at the endogenous locus [11]. The concept can be transferred to any target of interest. However, in the case of essential genes the coding sequence cannot be truncated. Further, fusion constructs need to be validated to be within the detection limit of the available equipment. The general workflow, calculation and corresponding data is visualized in Figure 7A.

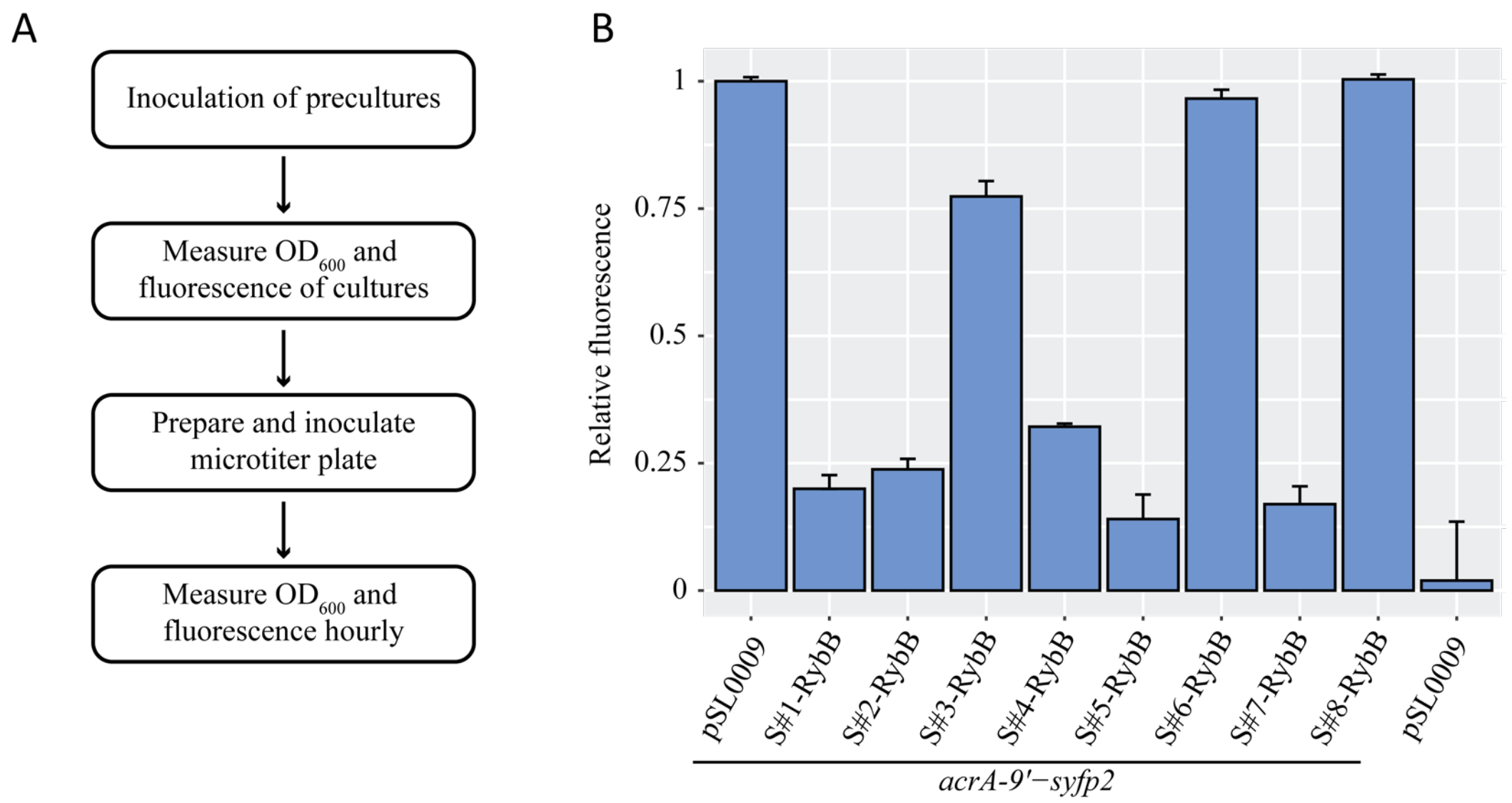

**Figure 7 | Functional testing of synthetic small RNAs with a fluorescence reporter.** (A) The workflow consists of the preparation and conduction of the experiment with subsequent data analysis. The protocol guides step by step through this process. (B) Measured fluorescence values for the constructed synthetic sRNAs in comparison to an empty plasmid control (pSL009) in the reporter (left) and wild type strain (right) after overnight incubation. For better comparison the empty plasmid control of the reporter strain was set to 1. Error bars represent the standard deviation. The results are consistent with the alternative benchmarking procedure in liquid media and the presence of oxacillin indicating a fluorescence reporter-based benchmarking may not be necessary. However, it becomes imperative if no phenotype can be evaluated.

1. Generate competent cells of desired reporter strain (here: *E. coli* MG1655 *acrA-9′−syfp2*) and transform previously extracted plasmid DNA to obtain respective strains.

   ***Important:*** Create corresponding controls with an empty plasmid in the reporter strain and wild type cells serving as positive and negative controls, respectively.

2. Isolate at least three single candidates for subsequent fluorescence reporter assay and perform measurements in technical triplicates.

3. Inoculate precultures of isolates and control strains in appropriate volume with LB medium containing the antibiotic for plasmid maintenance. Cultivate overnight at 37 °C. (*see* **Note 29**)

4. Prepare a 96-well microtiter plate with 150 µL LB medium containing the appropriate antibiotic and inoculate with overnight culture either 1:100 or preferably using a 96-pin replica-plating stamp or in case available the Singer Instruments Rotor HDA.

5. For data acquisition multiple options are available (*see* **Note 30**):

   (a) Incubate microtiter plate on a shaking incubator at 37 °C and 200 rpm. Take hourly measurements of optical density at 600 nm and fluorescence measurement according to the used fluorophore.
   (b) Measure absorbance and fluorescence directly in the plate reader using a kinetic program. (*see* **Note 31**)

6. Save output data as a csv file.

7. The optical density (OD) and fluorescence (sYFP2) values contain background intensities. The obtained data needs to be corrected. Subtract background values (typically values of the medium used for cultivation, here: LB) from optical density and fluorescence raw values:

```
sYFP2 of samples – sYFP2 of LB medium
OD600 of samples – OD600 of LB medium
```

8. After background correction, fluorescence intensity should be normalized to the optical density of a bacterial culture. Divide the fluorescence intensity by the optical density.

```
(sYFP2 of samples – sYFP2 of LB medium) / (OD600 of samples – OD600 of LB medium)
```

9. Generate plot for sRNA functionality assessment (Fig. 7B). (*see* **Note 32**)

### *3.6 Summary & Perspectives*

The stepwise protocol outlines the *in silico* design, *in vitro* construction and *in vivo* characterization of synthetic sRNAs. The predicted and characterized constructs show that a careful validation of synthetic sRNAs is important prior to their application. Based on our analysis, seed sequences S#1, S#4 and S#5 function superior to the other predicted seed sequences when RybB is used as sRNA scaffold. Notably, this protocol is the first time SEEDling predictions are used to create functional synthetic sRNAs. In the presented examples, 16-nucleotide seed sequences were used based on the natural seed region length of RybB. However, the detailed procedure of this protocol allows for the systematic testing to optimize synthetic sRNAs. The protocol can be adjusted in regard to the target mRNA, the seed region and/or the sRNA scaffold. Further, the system can be used with any available Golden Gate compatible cloning system by only adjusting the used type IIs enzyme and corresponding overhangs.

## 4. Notes

1. Check the current SEEDling documentation for the recommended Docker version.

2. All enzymes can be stored at -20 °C and should be kept on ice while used to generate reactions. However, if other enzymes are used storage conditions may differ and the user should act accordingly.

3. Make sure to double check the SEEDling documentation which may have altered/improved in comparison to the here described version. Alternatively to the described way the image can also be obtained directly from Docker Hub:

```
docker pull cedkb/digger_bac-seedling
```

4. Annotated genomes in GenBank format for many organisms can be retrieved from the NCBI Genome database (https://www.ncbi.nlm.nih.gov/genome/).

5. Make sure all relevant type IIs recognition sites are excluded.

6. Pick your target genes or leave file empty for genome wide predictions.

7. Always run programs as administrator.

8. If no seed is given in the output, try adjusting the blast_evalue (especially for long seed regions).

9. Make sure the overhangs match your cloning system. This protocol is based on a pBAD derivative (pSL137) and the plasmids pSL135 and pSL123 containing the $P_L$lacO-1 promoter and *rybB* scaffold, respectively, and uses SapI as type IIs restriction endonuclease.

10. The concept is transferable to any other Golden Gate cloning system. However, SapI produces 3-nucleotides fusion sites and, therefore, reduces the number of unwanted bases to a minimum. Furthermore, the protocol can also be adapted to other promoters or scaffolds. They can be either generated by (i) annealed oligonucleotides, (ii) gene synthesis or (iii) as subcloned DNA fragments in a corresponding level 0 plasmid. To learn more about the concept of level 0 plasmids, we refer the reader to [17].

11. Prevent oligonucleotides from heat to avoid denaturation.

12. Absorption at 260 nm is usually sufficient (e.g., Nanodrop), but specific DNA quantification (e.g., by using a Qubit) might be preferable.

13. Multiple annealed oligonucleotides can be used as well. However, in this case the oligonucleotides need to be phosphorylated.

14. It is recommended to add the enzymes as the last components to the reaction mixture. Always make sure the enzymes are stored appropriately and kept on ice while handling.

15. Estimation of molarity: 50 ng of a 3-kb plasmid is approx. 25 fmol.

16. Remaining reaction mix can be stored at -20 °C. In case no colonies are obtained after transformation, the whole mix can be transformed. Trouble shooting if no candidates are obtained:
    (1) check if overhangs are correct
    (2) check DNA parts on an agarose gel if band intensity matches measured DNA concentration
    (3) determine competence of cells by transforming 1 ng of a standard plasmid
    (4) check if the correct antibiotic was used and the recovery phase was appropriate
    (5) check if used enzymes are functional with a test digest of a known plasmid

17. Keep tubes in the fridge to inoculate positive candidates for cryo-cultures and plasmid extraction.

18. Thiazole orange is considered as a safe alternative for ethidium bromide and can be detected under blue light. 10,000 x stock = 13 mg/mL Thiazole orange in DMSO [18].

19. Generally 1% agarose concentration is good. However, if the fragments are < 500 bps, it can be helpful to increase the agarose concentration for better resolution.

20. In this protocol an open-source protocol based on magnetic bead purification using SeraMag SpeedBeads was used. Detailed information can be found on www.bomb.bio [19].

21. Our previous study showed that sRNAs regulation varies [11]. For the regulation of the *acrA* gene in combination with the RybB scaffold, concentrations of 25, 50, 75 and 100 µg/mL oxacillin are recommended.

22. Air bubbles can be removed or at least moved to the edges of the agar plate by using a sterile tip.

23. Use of 12-channel multichannel pipettes increases accuracy and speed of the experiment. Make sure to properly mix each dilution by pipetting and use new tips for each dilution step.

24. Make sure that the plates are properly dried. Placing the plates to 37 °C prior spotting can improve the drying of the spots. If drops fuse, reduce the amount of liquid transferred. Placing a printed 96-well grit below the plate can improve the manual spotting.
25. Besides the general colony numbers in the dilution steps, a reduction of the colony diameter can be a hint for a growth phenotype.
26. Cultures can be used to generate a working plate for subsequent experiments allowing to inoculate precultures in 96-well format using 96-pin based replica plating tools or a Singer Instruments Rotor HDA+. Cryo-cultures are generated by combining 70 µL of overnight culture with 30 µL 50% glycerol. Make sure to properly mix before sealing with an aluminium seal and store the sealed 96-well plates at -70 °C.
27. It is recommended to stick to a pattern. For example, for up to 12 candidates the top half could be the condition without compound and each row would be one biological replicate. The columns would be the respective candidates. The lower half of the plate could be the condition with compound. In this scenario the measurement for +/- condition and each biological replicate would be identical.
28. In our setting we used a CLARIOstar Plus plate reader with a particular plate holder for long term kinetics under shaking conditions with the following settings: 37 °C, orbital shaking during idle time, 120 seconds 600 rpm linear shaking before absorbance measurement at 600 nm with a total cycle time of 300 seconds. Measurements were taken for up to 24 hours.
29. Cultivation directly into 96-well microtiter plates is recommended and allows for a first measurement after overnight incubation.
30. For measurements > 6 hours sealing of microtiter plates is recommended to avoid evaporation. Make sure that the seal is optically clear and allows oxygen diffusion for proper fluorophore maturation.
31. It is recommended to characterize the fluorescence reporter initially in a kinetic to identify the point where the best dynamic range is expected.
32. In case of kinetic analysis, use normalized optical density and fluorescence values to plot a graph containing a secondary y-axis for the comparison of growth and fluorescence.

**Acknowledgement and Statement**

This work was supported by the Max Planck Society within the framework of the MaxGENESYS project (DS), the European Union (NextGenerationEU) via the European Regional Development Fund (ERDF) by the state Hesse within the project "*biotechnological production of reactive peptides from waste streams as lead structures for drug development*" (DS), and a Boehringer Exploration Grant (BAB). We are grateful to all laboratory members for extensive discussions on synthetic sRNAs in particular Cedric Brinkman for the development of SEEDling. All material is available from the corresponding author upon request.